\documentstyle[12pt]{article}
  \newcommand{\blankline}{\vskip .3cm}
  \newcommand{\f}{\begin{equation}}
  \newcommand{\ff}{\end{equation}}
\begin{document}
  \centerline{\Large  {\bf  Quantum symmetry, the cosmological constant}}
  \centerline{\Large {\bf  and Planck scale phenomenology}}
  \blankline
  \blankline
  \blankline
  \rm
  \centerline{{\bf Giovanni~AMELINO-CAMELIA}$^{a,b}$,{\bf Lee SMOLIN}$^{b,c}$,
  {\bf Artem STARODUBTSEV}$^{b,c}$}
  \blankline
  \centerline{\it $^a$Dipart.~Fisica,
Univ.~Roma ``La Sapienza'', and INFN Sez.~Roma1}
  \centerline{\it P.le Moro 2, 00185 Roma, Italy}
    \centerline{\it $^b$Perimeter Institute for Theoretical Physics,
    Waterloo,  Canada}
\centerline{\it $^c$Department of Physics, University of Waterloo,
Waterloo,  Canada}
%  \centerline{lsmolin@perimeterinstitute.ca}
  \blankline
  \blankline
  \blankline
  \blankline
  \blankline
  \blankline
  \blankline

\centerline{ABSTRACT}
We present a simple algebraic argument for the conclusion that the
low energy limit of a quantum theory of gravity must be a theory
invariant, not under the Poincar\'{e} group, but under a deformation
of it parameterized by a dimensional parameter proportional
to the Planck mass. Such deformations, called $\kappa$-Poincar\'{e}
algebras, imply modified energy-momentum relations of a type
that may be observable in near future experiments.  Our argument
applies in both $2+1$ and $3+1$ dimensions and
assumes only 1) that the low energy limit of a quantum theory
of gravity must involve also a limit in which the
cosmological constant is taken very small with respect to the Planck scale
and 2) that in $3+1$ dimensions the physical energy and momenta of
physical
elementary particles is related to symmetries of the full quantum
gravity theory by appropriate renormalization depending on
$\Lambda l^2_{Planck}$.  The argument makes use of the fact that
the cosmological constant results in the symmetry algebra of quantum
gravity being quantum deformed, as a consequence when the
limit $\Lambda l^2_{Planck} \rightarrow 0$ is taken one finds a
deformed Poincar\'{e} invariance.  We are also able to isolate
what information must be provided by the quantum theory in order
to determine which presentation of the $\kappa$-Poincar\'{e} algebra
is relevant for the physical symmetry generators and, hence, the
exact form of the modified energy-momentum relations.
These arguments imply that Lorentz invariance is modified
as in proposals for doubly special relativity, rather than broken, in
theories of quantum gravity, so long as those theories behave smoothly in
the limit the cosmological constant is taken to be small.

\vfill
\eject

\section{Introduction}

The most basic questions about quantum gravity concern the nature of
the fundamental
length, $l_p= \sqrt{\hbar G/c^3}$. One possibility, which has been explored
recently by many authors (see Refs.~\cite{qgprev} and references therein),
is that it acts as a threshold for
new physics, among
which is the possibility of deformed energy-momentum relations,
\f
E^2 =p^2 + m^2 + \alpha l_p E^3 + \beta l_p^2 E^4 + \ldots
\label{deformed}
\ff
As has been discussed in many places,
this has consequences for present and near
term observations~\cite{qgprev}.
However, when analyzing the phenomenological consequences of (\ref{deformed}),
there are two very different possibilities which must be
distinguished. The first is that the relativity of inertial frames
no longer holds, and there is a preferred frame. The second is that
the relativity
of inertial frames is maintained but, when comparing measurements made
in different frames, energy and momentum must be transformed non-linearly.
This latter possibility, proposed in Ref.~\cite{dsr1},
is called deformed or doubly special relativity (DSR)~\footnote{Various
consequences of and issues concerning this hypothesis are discussed
in Ref.~\cite{dsrpaps,various}
and references therein.}.

Both modified energy momentum relations, (\ref{deformed}) and DSR should,
if true,
be consequences of a fundamental quantum theory of gravity.  Indeed, there
are
calculations in loop quantum gravity\cite{gampulmexweave,positive}
and other approaches\cite{grbgac,jacob} that give
rise to relations of type (\ref{deformed}). However, it has not been so far
possible to distinguish between the two possibilities of a
preferred quantum gravity frame and DSR.  Some calculations
that lead to (\ref{deformed}), such as \cite{gampulmexweave}, may be
described as studies of perturbations of weave states,
which themselves appear to pick out a preferred frame.
Further these states are generally non-dynamical in that
they are not solutions to the full set of constraints of
quantum gravity and there is no evidence they minimize a
hamiltonian. That there are some states of the theory whose
excitations have a modified spectrum of the form of
(\ref{deformed}) is not surprising, the physical question is whether
the ground state is one of these, and what symmetries it has.

In Ref.~\cite{positive}, one of us tried to approach the question of deformed
dispersion relations by deriving one, for the simple case of a
scalar field, from a state which is both an exact solution to all the
quantum constraints of quantum gravity and has, at least naively, the full
set of symmetries expected of the ground state. This is the
Kodama state\cite{kodama},
which requires that the cosmological constant be non-zero.  The result
was that scalar field excitations of the state do satisfy deformed
dispersion relations, in the limit that the cosmological constant
$\Lambda$ is taken to zero, when the effective field theory for the
matter field is derived from the quantum
gravity theory by a suitable renormalization of operators, achieved by
multiplication of a suitable power of $\sqrt{\Lambda} l_p$.

The calculation leading to this result was, however rather complicated, so
that one should wonder whether it is an accident or reflects an underlying
mathematical relationship. Furthermore, one may hope to isolate the
information that a quantum theory of gravity should provide to
determine the energy momentum relations that emerge for elementary
particles in the limit of low energies.
The purpose of this paper is to suggest that
there is indeed a deep reason for a $DSR$ theory to emerge from a quantum
gravity theory, when the latter has a non-zero bare cosmological constant,
and the definition of the effective field theory that governs the low-energy
flat-spacetime
physics involves a limit in which $\Lambda l_p^2 \rightarrow 0$. Our argument
has the following steps.

\begin{enumerate}

\item{}We first argue that even if the renormalized, physical,
cosmological constant vanishes, or is very small in Planck units,
it is still  the case that in any non-perturbative background
independent approach to quantum gravity, the parameters of the theory
should include a bare cosmological constant. It must be there in ordinary
perturbative approaches, in order to cancel contributions to the vacuum
energy coming from quantum fluctuations of the matter fields. A nonzero,
and in fact positive, bare $\Lambda$ is also required in non-perturbative,
background independent approaches to quantum gravity, such as dynamical
triangulations~\cite{triangles} or Regge calculus~\cite{reggecalc},
otherwise the theory has
no critical behavior required for a good low energy limit. There is also
evidence from loop quantum gravity that a nonzero bare $\Lambda$ is at least
very helpful, if not required, for a good low energy
limit\cite{positive,kodama}.  To extract the
low energy behavior of a quantum theory of gravity,
it will then be necessary to study the limit
$\Lambda l_p^2 \rightarrow 0$.

\item{}We then note that when there is a positive cosmological constant,
$\Lambda$, excitations of the ground state of a quantum gravity theory are
expected to transform under representations of the quantum deformed
deSitter algebra,
with
$z=\ln q$ behaving in the limit of small $\Lambda l_p^2$ as,
\begin{eqnarray}
z & \approx & \sqrt{\Lambda}l_p  \ \ \ \ \ \mbox{for} \ \ d=2+1
\ \ \ \cite{witten,regge,NR1}
\label{2+1}
\\
z & \approx & \Lambda l_p^2  \ \ \ \ \ \mbox{for} \ \ d=3+1  \ \
\cite{linking,baez-deform,qdef,artem}
\label{3+1}
\end{eqnarray}
Below we will summarize the evidence for this expectation.

\item{}For $d=2+1$ we note that the limit in which
$\Lambda l_p^2 \rightarrow 0$ then involves the simultaneous
limit $z \approx  \sqrt{\Lambda}l_p \rightarrow 0$. We note that
this contraction of $SO_q(3,1)$, which is the quantum deformed deSitter
algebra in $2+1$ dimensions, is not the classical Poincar\'{e}
algebra ${\cal P}(2+1)$,
as would be the case if $q=1$ throughout. Instead,
the contraction
leads to a modified $\kappa$-Poincar\'{e}~\cite{lukieAnnPhys},
algebra ${\cal P}_{\kappa}(2+1)$,
with the dimensional parameter $\kappa \approx l_p^{-1}$.
It is well known that some of these algebras provide the basis for
a DSR theory with a modified
dispersion relation of the form (\ref{deformed}).

\item{}For $d=3+1$ we note that the contraction
$\Lambda l_p^2 \rightarrow 0$ must be done scaling $q$
according to (\ref{3+1}). At the same time, the contraction
must be accompanied by the
simultaneous renormalization of the generators for energy and momentum of
the excitations, of the form
\f
E_{ren} = E \left ( {\sqrt{\Lambda} l_p \over \alpha } \right )^r, \ \ \
P^i_{ren} = P^i \left ( {\sqrt{\Lambda} l_p \over \alpha} \right )^r
\label{renormalize}
\ff
where $E_{ren}$ is
the renormalized energy
relevant for the effective field theory description and $E$ is
the bare generator from the quantum gravity theory.  (The power
$r$ and constant $\alpha$ must be the same in both cases to preserve the
 $\kappa$-Poincar\'{e} algebra.) This is expected because, unlike the case
in $2+1$, in $3+1$ dimensions there are local degrees of freedom,
whose effect on the operators of the effective field theory must be
taken into account when taking the contraction.

We then find that when $r=1$ the contraction is again
the  $\kappa$-Poincar\'{e} algebra, with $\kappa^{-1} = \alpha l_p$.
However when $r>1$ there is no good contraction, whereas when $r<1$ the
contraction is the ordinary Poincar\'{e}
 symmetry.  This was found also
explicitly for the case of a scalar field in Ref.~\cite{positive}.

 \item{}This argument assures us that whenever $r=1$ the symmetry
 of the ground state in the limit $\Lambda l_p^2 \rightarrow 0$
 will be $\kappa$ deformed Poincar\'{e}. However,
 there remains a freedom in the specification of the presentation
of the algebra relevant for the physical low energy operators that
generate translations in time and space, rotations and boosts, due to
the possibility of making non-linear redefinitions of the generators
of $\kappa$-Poincar\'{e}. Some of the freedom is tied down by requiring
that the algebra have an ordinary Lorentz subalgebra, which is
necessary so that transformations between measurements made by
macroscopic inertial observers can be represented. The remaining
freedom has to do with the exact definition of the energy and momentum
generators of the low energy excitations, as functions of operators
in the full non-linear theory.  As a result, the algebraic information
is insufficient to predict the exact form of the energy-momentum
relation, however it allows us to isolate what remaining information
must be supplied by the theory to determine them.

\end{enumerate}

Hence, for the unphysical case of $2+1$ dimensions
we then argue that so long as the low
energy behavior is defined through a limit $\Lambda l_p^2 \rightarrow 0$,
there is a very general argument, involving only symmetries, that tells us
that that limit is characaterized by low energy excitations transforming
under representations of the $\kappa$-Poincar\'{e} algebra. This means that the
physics is a DSR theory with deformed energy momentum relations
(\ref{deformed}), but with relativity of inertial frames preserved.

In the physical case of $3+1$ dimensions we conclude that the same is true, so
long as an additional condition holds, which is that the derivation
of the low energy theory involves a renormalization of the energy
and momentum generators of the form of (\ref{renormalize}) with $r=1$.

Hence we conclude that there is a very general algebraic structure that
governs the deformations of the energy-momentum relations at the Planck scale,
in quantum theories of gravity where the limit
$\Lambda l_p^2 \rightarrow 0$ is smooth.

Sections 2 and 3 are devoted, respectively, to the cases of $2+1$ and $3+1$
dimensions. One argument for the quantum deformation of the
algebra of observables in the $2+1$ dimensional case is reviewed in
the appendix.  Section 3 relies on results
on observables in $3+1$ dimensional quantum gravity by one of
us\cite{artem}.

\section{DSR symmetries in $2+1$ Quantum Gravity}

\subsection{$2+1$ Quantum Gravity}

In this subsection we will review the basics of quantization of
general relativity in $2+1$ dimensions and how quantum groups come
into play.

$2+1$ quantum gravity has been a subject of extensive study since
mid 80's and now this is a well understood theory (at least in the
simple case when there is no continuous matter sources). For
nonzero cosmological constant the theory is described by the
following first order action principle
\begin{equation}
S=-\frac{1}{2G} \int \big( d \omega^{ij}-\omega^i_k \wedge
\omega^{kj} - \Lambda e^i \wedge e^j \big)\wedge e^l
\epsilon_{ijl}, \label{action1}
\end{equation}
where $\omega^{ij}$ is an SO(2,1)-connection 1-form, $e^i$ is a triad
1-form, $i,j=0,1,2$ and G is the Newton constant. The equations of
motion following from the action (\ref{action1})
\begin{equation}
d \omega^{ij}-\omega^i_k \wedge \omega^{kj} - 3\Lambda e^i \wedge
e^j=0 \label{dscurv}
\end{equation}
simply mean that the curvature is constant everywhere and
therefore  geometry doesn't have local degrees of freedom. The
theory is not completely trivial however. One can introduce so
called topological degrees of freedom by choosing a spacetime
manifold which is not simply connected.  Coupling to point particles
may be accomplished by adding extrinsic
delta-function
sources of curvature which represent point particles~\cite{hooft}.

Below we consider
spacetime  ${\cal M}$ to be  ${\cal M}=\Sigma^2 {\times} R^1$,
where $\Sigma^2$ is a compact spacelike surface of genus $g$.

The easiest way to solve 2+1 gravity is through rewriting the
action (\ref{action1}) as a Chern-Simons action for $G= SO(2,2)$ or
$SO(3,1)$ , depending on the sign of the cosmological constant
\cite{witten}, which reads
\begin{equation}
S=\frac{k}{4\pi} \int\big( d A^{ab}-\frac{2}{3}A^a_t \wedge
A^{tb} \big)\wedge A^{cd} \epsilon_{abcd}, \label{action2}
\end{equation}
where $a,b=0,1,2,3$, and $k=(l_p\sqrt{\Lambda})^{-1}$
is the
dimensionless coupling constant of the Chern-Simons theory.
(\ref{action1}) is obtained by decomposing say SO(2,2)-connection
$A^{ab}$ as $A^{ij}=\omega^{ij}$ and $A^{i3}=\sqrt{\Lambda}e^i$.
The action (\ref{action2}) leads to the standard canonical
commutation relations
\begin{equation}
[A^{ab}_\alpha(x),A^{cd}_\beta(y)]={2\pi \over k} \epsilon_{\alpha
\beta}\epsilon^{abcd} \delta^2(x,y) \label{ccr}
\end{equation}
and equations of motion saying that the connection $A^{ab}$ is
flat.

By now it is well understood that the resulting theory is described
in terms of the quantum deformed deSitter algebra, $SO_q(3,1)$
with $q$ given by $q=e^{2\pi \imath / k+2}$~\cite{witten,regge,NR1}.
For interested readers we review one approach to this conclusion,
which is given in the appendix.  However, the salient point is that
in the theory particles are identified with punctures in the
$2$ dimensional spatial manifold. These punctures are labeled by
representations of the quantum symmetry algebra $SO_q(3,1)$
\cite{NR1}.
However, in the limit of $\Lambda \rightarrow 0$ and of low
energies, these particles should be labeled by representations
of the symmetry algebra of the ground state. Furthermore, as there
are no local degrees of freedom in $2+1$ gravity there is no need
to study a limit of low energies, it should be sufficient to find
the symmetry group of the ground state in the limit
$\Lambda \rightarrow 0$ to ask what the contraction is of the
algebra whose representations label the punctures.
We now turn to that calculation.

\subsection{Contraction and  $\kappa$-Poincar\'{e} algebra in
$2+1$ dimensions}

In the previous subsection we have seen that in $2+1$ dimensional
quantum gravity with $\Lambda \neq 0$ a quantum symmetry
algebra $SO_q(3,1)$ or
$SO_q(2,2)$  arises as a result of canonical non-commutativity of
the Chern-Simons (anti)deSitter connection, and its representations label
the punctures that represent particles. . The canonical
commutation relations of the Chern-Simons theory determine the
deformation parameter $z$ to be linear in $\sqrt{\Lambda}$ as in (\ref{2+1}).
The next step in our argument is to describe the $\Lambda \rightarrow 0$
limit of 2+1dimensional
quantum gravity, focusing on the structure of the symmetry algebra
that replaces $SO_q(3,1)$ in the limit.

In order to rely on explicit formulas let us
adopt an explicit basis for $SO_q(3,1)$. We describe $SO_q(3,1)$
in terms of the six
generators $M_{0,1},M_{0,2},M_{0,3},M_{1,2},M_{1,3},M_{2,3}$
satisfying the commutation relations:
\begin{eqnarray}
&& [M_{2,3},M_{1,3}]  =  {1 \over z} \sinh(z M_{1,2}) \cosh(z M_{0,3})
\nonumber \\
&& [M_{2,3},M_{1,2}] = M_{1,3}
\nonumber \\
&& [M_{2,3},M_{0,3}] = M_{0,2}
\nonumber \\
&& [M_{2,3},M_{0,2}] =  {1 \over z} \sinh(z M_{0,3}) \cosh(z M_{1,2})
\nonumber \\
&& [M_{1,3},M_{1,2}] = - M_{2,3}
\nonumber \\
&& [M_{1,3},M_{0,3}] = M_{0,1}
\nonumber \\
&& [M_{1,3},M_{0,1}] =   {1 \over z} \sinh(z M_{0,3}) \cosh(z M_{1,2})
\nonumber \\
&& [M_{1,2},M_{0,2}] = - M_{0,1}
\nonumber \\
&& [M_{1,2},M_{0,1}] = M_{0,2}
\nonumber \\
&& [M_{0,3},M_{0,2}] = M_{2,3}
\nonumber \\
&& [M_{0,3},M_{0,1}] = M_{1,3}
\nonumber \\
&& [M_{0,2},M_{0,1}] =  {1 \over z} \sinh(z M_{1,2}) \cosh(z M_{0,3})
\label{so31q}
\end{eqnarray}
with all other commutators trivial.

The reader will easily verify that in the $z \rightarrow 0$ limit
these relations reproduce the $SO(3,1)$ commutation relations.
In addition, it is well known that upon setting $z=0$ from the beginning
(so that the algebra is the classical $SO(3,1)$) we should make the
identifications
\begin{eqnarray}
&& E = \sqrt{\Lambda} M_{0,3}
\nonumber \\
&& P_i = \sqrt{\Lambda} M_{0,i}
\nonumber \\
&& M = M_{1,2}
\nonumber \\
&& N_i = M_{i,3}
\label{ep}
\end{eqnarray}
%\begin{eqnarray}
%&& E = \sqrt{\Lambda} M_{0,1}
%\nonumber \\
%&& P_1 = \sqrt{\Lambda} M_{0,2}
%\nonumber \\
%&& P_2 = \sqrt{\Lambda} M_{0,3}
%\nonumber \\
%&& N_1 = M_{1,2}
%\nonumber \\
%&& N_2 = M_{1,3}
%\nonumber \\
%&& M = M_{2,3}
%\label{ep}
%\end{eqnarray}
This makes manifest the well known fact that
the Inonu-Wigner~\cite{inWig}
contraction  $\Lambda \rightarrow 0$ of the deSitter algebra
$SO(3,1)$ leads to the
classical Poincar\'{e} algebra ${\cal P}(2+1)$.

However, in quantum gravity, we cannot take first the classical limit
$z\rightarrow 0$ and then the contraction $\Lambda \rightarrow 0$ because,
by the relation (\ref{2+1}),
the two parameters are proportional to each other.
The limit must be taken so that the ratio
\f
\kappa^{-1} = {z \over \sqrt{\Lambda}} = l_p
\label{2+1ratio}
\ff
is fixed. The result is that the limit is not the classical
Poincar\'{e} algebra, it is instead
the  $\kappa$-Poincar\'{e}~\cite{lukieAnnPhys}
algebra ${\cal P}_{\kappa}(2+1)$. This is easy to see.
We rewrite (\ref{so31q})
using (\ref{ep}) and assuming  (\ref{2+1}) we find,
\begin{eqnarray}
&& [N_2,N_1]  = {1 \over z} \sinh(z M) \cosh(z E/\sqrt{\Lambda})
= {1 \over l_p \sqrt{\Lambda}} \sinh(l_p \sqrt{\Lambda} M)
\cosh(l_p E)
\nonumber \\
&& [M,N_i] = \epsilon_{ij} N^j
\nonumber \\
&& [N_i,E] = P_i
\nonumber \\
&& [N_i,P_j] = \delta_{ij}
\sqrt{\Lambda} {1 \over z} \sinh(z E/\sqrt{\Lambda}) \cosh(z M)
=\delta_{ij} {1 \over l_p} \sinh(l_p E) \cosh(l_p \sqrt{\Lambda} M)
\nonumber \\
&& [M,P_i] = \epsilon_{ij} P^j
\nonumber \\
&& [E,P_i] = \Lambda  N_i
\nonumber \\
&& [P_2,P_1] = \Lambda {1 \over z} \sinh(z M) \cosh(z E/\sqrt{\Lambda})
= {\sqrt{\Lambda} \over l_p } \sinh(L_p \sqrt{\Lambda} M)
\cosh(L_p E)
\label{so31qep}
\end{eqnarray}
From this it is easy to obtain the $\Lambda l_p^2 \rightarrow 0$ limit:
\begin{eqnarray}
&& [N_2,N_1]  =  M \cosh(l_p E)
\nonumber \\
&& [M,N_i] = \epsilon_{ij} N^j
\nonumber \\
&& [N_i,E] = P_i
\nonumber \\
&& [N_i,P_j] = \delta_{ij} {1 \over l_p} \sinh(l_p E)
\nonumber \\
&& [M,P_i] = \epsilon_{ij} P^j
\nonumber \\
&& [E,P_i] = 0
\nonumber \\
&& [P_2,P_1] = 0
\label{so31qPoinc}
\end{eqnarray}
which indeed assigns deformed commutation relations for the generators
of Poincar\'{e} transformations.
The careful reader can easily verify that,
upon imposing $\kappa = l_p^{-1}$, Eq.~(\ref{so31qPoinc}) gives
the commutation relations\footnote{We note that
the  ${\cal P}(2+1)_{\kappa}$ algebra can be endowed
with the full structure of a Hopf algebra (just like $SO_q(3,1)$),
and that it is known in the literature that the full Hopf algebra
of ${\cal P}(2+1)_{\kappa}$ can be derived from the quantum
group $SO_q (3,1)$ by contraction. However,
in this paper we focus only on the commutation relations needed
to make our physical argument.}
characteristic of
the ${\cal P}(2+1)_{\kappa}$ $\kappa$-Poincar\'{e} algebra
described in Ref.~\cite{lukieDdim}.

It is striking that the fact that the $z \rightarrow 0$ and
$\Lambda l_p^2  \rightarrow 0$ limit must be taken together,
for physical reasons, leaves us no alternative but
to obtain a
deformed  Poincar\'{e} algebra. This is because the {\it dimensional}
ratio (\ref{2+1ratio}) is fixed during the contraction, and
appears in the resulting algebra. No dimensional scale appears in the
classical Poincar\'{e} algebra, so the result of the contraction cannot
be that, it must be a deformed algebra labeled by
the scale $\kappa$\footnote{To our knowledge this is the first
example of a context in which
a Inonu-Wigner contraction takes one automatically from
a given quantum algebra to another quantum algebra.
Other examples of Inonu-Wigner contraction of
a given quantum algebra have been considered in the literature
(notably in Refs.~\cite{firenze,lukieIW}),
but in those instances there is no {\it a priori} justification
for keeping fixed the relevant ratio of
parameters, and so one has freedom to choose whether
the contracted algebra is classical or quantum-deformed.}.

At the same time, there is  freedom in defining the presentation of
the algebra that results in the limit. This is possible because we can scale
various of the generators as we take the contraction. For physical
reasons we want to exploit this. A problem with the presentation
just given in (\ref{so31qPoinc}) is that the generators of the
$SO(2,1)$ Lorentz algebra do not close on the usual Lorentz algebra,
hence they do not generate an ordinary transformation group. However
if the generators of boosts and rotations are to be interpreted
physically as giving us rules to transform measurements made by
different macroscopic intertial observers into each other, they must
exponentiate to a group, because the group properties follow directly
from the physical principle of equivalence of inertial frames.

We would then like to choose a different presentation of
the $\kappa$-Poincar\'{e} algebra in which the lorentz generators
form an ordinary Lie algebra.
There is in fact more than one way to accomplish this, because there
is freedom to scale all the generators as the contraction
is taken by functions of $l_pE$.  The identification of the correct
algebra depends on additional physical information about how the generators
must scale as the limit $\Lambda l_p^2 \rightarrow 0$ is taken.

In the absence of additional physical input, we give here one
example of scaling to a presentation which contains an undeformed
Lorentz algebra. It  is defined by
replacing (\ref{ep}) by the following
definitions of the energy, momenta, and boosts~\cite{MajidRuegg},
\begin{eqnarray}
&& E = \sqrt{\Lambda} M_{0,3}
~ , ~~ \exp[z E/(2 \sqrt{\Lambda})] P_1 = \sqrt{\Lambda} M_{0,1}
~ , ~~ \exp[z E/(2 \sqrt{\Lambda})] P_2 = \sqrt{\Lambda} M_{0,2}
\nonumber \\
&& \exp[z E/(2 \sqrt{\Lambda})] \left( N_1
- {z \over 2 \sqrt{\Lambda}} M P_2  \right) = M_{1,3}
~ , ~~ M = M_{1,2}
\nonumber \\
&& \exp[z E/(2 \sqrt{\Lambda})] \left( N_2
+ {z \over 2 \sqrt{\Lambda}} M P_1 \right) = M_{2,3} ~.
\label{epbis}
\end{eqnarray}
We again take the contraction keeping the ratio (\ref{2+1ratio}) fixed.
Then the ${\cal P}(2+1)_{\kappa}$ commutation relations obtained
in the limit $\Lambda l_p^2 \rightarrow 0$ take the
following form~\cite{lukieAnnPhys,MajidRuegg}.

\begin{eqnarray}
&& [N_2,N_1]  =  M
\nonumber \\
&& [M,N_i] = \epsilon_{ij} N^j
\nonumber \\
&& [N_i,E] = P_i
\nonumber \\
&& [N_2,P_2] = - {1 - e^{2 l_p E} \over 2 l_p} - {l_p \over 2} P_1^2
+ {l_p \over 2} P_2^2
\nonumber \\
&& [N_1,P_1] =  - {1 - e^{2 l_p E} \over 2 l_p} - {l_p \over 2} P_2^2
+ {l_p \over 2} P_1^2
\nonumber \\
&& [M,P_i] = \epsilon_{ij} P^j
\nonumber \\
&& [E,P_i] = 0
\nonumber \\
&& [P_1,P_2] = 0 ~.
\label{so31qPoincbis}
\end{eqnarray}

In the literature this is called the  bicross-product
basis ~\cite{lukieAnnPhys,MajidRuegg}.

We see that the Lorentz generators form a Lie algebra, but the
generators of momentum transform non-linearly. This is characteristic
of a class of theories called, {\it deformed} or {\it doubly
special relativity theories}~\cite{dsr1,dsrpaps,various}, which have
recently been studied in the
literature from a variety of different points of view. The main idea
is that the relativity of inertial frames is preserved,
but the laws of transformation between different frames are now
characterized by two invariants, $c$ and $\kappa$ (rather than the single
invariant c of ordinary special-relativity transformations).
This is possible because the momentum
generators transform non-linearly under boosts.  In fact, the
presentation just given was the earliest form of such a theory to be
proposed~\cite{dsr1}.

One consequence of the fact that the momenta transform non-linearly
under boosts is that the energy-momentum relations are
modified because the invariant function of $E$ and $P_i$ preserved
by the action described above in (\ref{so31qPoincbis}) is no longer
quadratic.
Instead, the (dimensionless) invariant mass is given by\footnote{We can also
express the invariant mass $M$ in terms of the rest
energy $m$ by $M^2 =\cosh(L_p m)$.}
\begin{eqnarray}
M^2 \equiv  \cosh(l_p E)-\frac{l_p^2}{2}\vec{P}^2 e^{L_p E}
~,
\label{casimir}
\end{eqnarray}
This gives corrections
to the dispersion relations which are only linearly suppressed
by the smallness of $l_p$, and are therefore, as recently
established~\cite{qgprev,grbgac}, testable with the sensitivity
of planned observatories.

To conclude, we have found that the
limit $\Lambda l_p^2 \rightarrow 0$ of $2+1$
dimensional quantum gravity must lead to a theory where
the symmetry of the ground state is the $\kappa$-Poincar\'{e} algebra.
Which form of that algebra governs the transformations of
physical energy and momenta, and hence the exact deformed
energy-momentum relations, depends on additional physical information.
This is needed to fix the form of the low energy symmetry generators
in terms of the generators of the fundamental theory.

\section{The case of  $3+1$ Quantum Gravity}

Now we discuss the same argument in the case of $3+1$ dimensions.

\subsection{The role of the quantum deSitter algebra in $3+1$ quantum gravity}

The algebra that is relevant for the transformation properties of
elementary particles is the symmetry algebra. In classical or quantum
gravity
in $3+1$ or more dimensions, where there are local degrees of freedom,
this cannot be computed from symmetries of a background spacetime
because there is no background spacetime. Nor is this necessarily the
same as the algebra of local gauge transformations.  So we have to ask
the question of how, in quantum gravity, we identify the generators
of operators that will, in the weak coupling limit, become the
generators of transformations in time and space? That is,
how do we identify the operators that, in the low energy
limit in which the theory is dominated by excitations of a state
which approximates a maximally symmetric spacetime,  become the
energy, momenta, and angular momenta?

The only answer we are aware of which leads to results is to impose a
boundary, with suitable boundary conditions that allow symmetry
generators to be identified as operations on the boundary. In fact
we know that in general relativity the hamiltonian, momentum and
angular momentum operators are defined in general only as boundary
integrals. Further they are only meaningful when certain boundary
conditions have been imposed. A necessary condition for energy
and momenta to be defined is that the lapse and shift are fixed,
then the energy and momenta can be defined as generators parameterized
by the lapse and shift.

In seeking to define energy and momentum, we can make use of a set
of results which have shown that in both
the classical and quantum theory boundary conditions can be imposed in
such a way that the full background independent dynamics of the bulk
degrees of freedom can be studied~\cite{linking}. There are further results
on boundary Hilbert spaces and observables which show that
physics can indeed be extracted in quantum gravity from studies
of theories with boundary conditions, such as the studies of black
hole and cosmological horizons~\cite{leeNO2}.

We argue here that this method can be used to extract the exact
quantum deformations of the boundary observables algebra, and that
the information gained is sufficient to repeat the argument just
given in one higher dimension.  More details
on this point are given in a paper
by one of us~\cite{artem}.

In fact it has already been shown that the boundary observables algebras
relevant for $3+1$ dimensional quantum gravity become quantum deformed
when the cosmological constant is turned on, with
\f
q=e^{2\pi\imath / k +2 }
\ff
with the level $k$ given by\cite{linking,baez-deform,positive,qdef}
\f
k={6 (\imath) \pi \over G\Lambda  }
\label{kvalue}
\ff
where the $\imath$ is present in the case of the Lorentzian theory
and absent in the case of the Euclidean theory.
This gives (\ref{3+1}) in the limit of small cosmological constant.
These stem from the observation that classical gravity theories,
including, general relativity and supergravity (at least up to $N=2$)
can be written as deformed topological field theories, so that their actions
are of the form of
\f
S^{bulk}= \int_{\cal M}  \left [ Tr
( B \wedge F - {\Lambda \over 2} B \wedge B ) -
Q( B \wedge B )
\right ]
\ff
Here $B$ is a two form valued in a lie algebra $G$, $F$ is the curvature
of a connecton, $A$, valued in $G$ and $ Q( B \wedge B )$ is a quadratic
function of the components\footnote{This form holds in all dimensions,
see \cite{leeNO3},
it also extends to supergravity with $G$ a superalgebra\cite{leeNO4}.}.
Were the last term absent, this would be a topological field theory.

In the presence of a boundary, one has to add a boundary term to the
action and
impose a boundary condition. One natural boundary condition, which has
been much studied, for a
theory of this form is
\f
F= \Lambda B
\ff
pulled back into the boundary.   The resulting boundary term is
the Chern-Simons
action of $A$ pulled back into the boundary,
\f
S=S^{bulk} + S^{boundary}, \ \ \ S^{boundary}=
{k \over 4\pi} \int_{\partial {\cal M}}Y(A)_{CS}
\ff
where $Y(A)_{CS} = Tr (A \wedge dA + {2\over 3} A^3 )$ is the
Chern-Simons three form. Consistency with the equations of motion
then requires that (\ref{kvalue}) be imposed.  This leads to a quantum
deformation of the algebra whose representations label spin networks and
spin foams, as shown in Ref.~\cite{linking,baez-deform}.
It further leads to a quantum
deformation of the algebra of observables on the boundary\cite{linking}.

In $3+1$ dimensions there are several choices for the group $G$,
that all lead to
theories that are classically equivalent to general relativity
(for non-degenerate
solutions).  One may take $G=SU(2)$, in which case $Q(B\wedge B)$
can be chosen so
that the bulk action, $S^{bulk}$ is the Plebanski action and the corresponding
hamiltonian formalism is that of Ashtekar\cite{leeNO7}.
In this case the addition of
the cosmological constant and boundary term leads to spin networks labeled by
$SU_q (2)$, with (\ref{kvalue}). One can also take
$G=SO(3,1)$ and choose $Q(B\wedge B)$ so that $S^{bulk}$ is the Palatini action.
From there one can derive a spin foam model, for example the Barrett-Crane
model\cite{leeNO8}.
By turning on the cosmological constant, one gets the Noui-Roche spin foam
model\cite{NR2},
based on the quantum deformed lorentz group $SO_q (3,1)$, with $q$ given
still by
(\ref{kvalue}).

However, as shown in \cite{leeNO9,artem},in the case of
non-zero $\Lambda$ we can also choose $G$ to be the (A)dS group,
$SO (3,2)$ or $SO(4,1)$.
In this case, as shown in \cite{artem} one can also study a
different boundary condition, in which the metric pulled back
to the boundary is fixed. This has the advantage that it allows
momentum and energy to be defined on the boundary, as lapse and
shift can be fixed. In this case we can take the boundary action
to be the Chern-Simons invariant of the (anti)deSitter algebra
group, with the $SO(4,1)/SO(3,1)$ coset labeling the frame
fields~\cite{artem}.

The resulting algebra of boundary observables is studied  in \cite{artem},
where it is shown that the boundary observables
algebra includes the subgroup
of the global $3+1$ deSitter group that leaves the boundary fixed.
For $\Lambda >0$ this is  $SO_q (3,1)$, with $q$ given again by
(\ref{kvalue}). Furthermore, the operators which generate global
time translations, as well as translations, rotations and boosts that
leave the boundary fixed can be identified, giving us a physically
prefered basis for the quantum algebra $SO_q (3,1)$.

This tells us that, were the geometry in the
interior frozen to be the spacetime with maximal symmetry,
the full symmetry group
must be $SO_q(4,1)$ with the same $q$\footnote{Another argument for
the relevence of the quantum deformed
deSitter group comes from recent work in spin foam models.
Several recent papers on spin foam models argue for a model based,
for $\Lambda=0$, on the representation theory of the
Poincar\'{e}  group\cite{CY}.
This fits nicely into a $2$-category framework\cite{CY}. When
$\Lambda >0$ one
would then replace the Poincar\'{e} group by the deSitter group, but
agreement with the
the previously mentioned results would require it be quantum deformed,
so we arrive
at a theory based on the representations of  $SO_q (3,2)$.

Yet another argument leading to the same conclusion comes from the
existence of
the Kodama state\cite{kodama,positive}, which is an exact physical quantum state
of the gravitational
field for nonzero $\Lambda$, which has a semiclassical interpretation
in terms of
deSitter. One can argue that a large class of gauge and diffeomophism
invariant
perturbations of the Kodama state are labeled by quantum spin networks
of the algebra
$SU_q(2)$ with again (\ref{kvalue})\cite{positive}.  However, of those,
there should be a
subset which describe gravitons with
wavelengths $\sqrt{\Lambda} > k > E_{Planck}$,
moving on the deSitter background as such states are known to exist
in a semiclassical
expansion around the Kodama state\cite{positive}. One way to construct such
states is to
construct quantum spin network states for $SO_q(3,2)$, and
decompose them into sums of
quantum spin network states for $SU_q(2)$. The different
states will then be labeled
by functions on the coset $SO_q(3,2)/SU_q(2)$.}.

\subsection{Contraction of the quantum deSitter algebra in $3+1$.}

We now study the contraction of the quantum deformed deSitter algebra in
$3+1$ dimensions. We first give a general argument, then we discuss the
boundary observables algebra of Ref.~\cite{artem}.

Our general argument is based on the observations reported in the previous
subsection concerning the role of the quantum algebras $SO_q(4,1)$
and $SO_q(3,2)$, with $\ln q \sim \Lambda l_p^2$ (for small $\Lambda$),
in quantum gravity in 3+1 dimensions.

Now, it is in fact known already in the literature
~\cite{lukieIW} that the $\Lambda \rightarrow 0$
contraction of these quantum algebras can lead
to the $\kappa$-deformed Poincar\'{e} algebra ${\cal P}_{\kappa}(3+1)$,
if the  $\Lambda \rightarrow 0$ is combined
with an appropriate $\ln q \rightarrow 0$ limit.
The calculations are rather involved, and are already discussed
in detail in  Ref.~\cite{lukieIW}. Hence, for our purposes here
it will be enough to focus on how the limit goes for
one representative $SO_q(3,2)$ commutator.
What we want to show is that quantum gravity in 3+1 dimensions
has the structure of the $\ln q \rightarrow 0$ limit,
which is associated to the $\Lambda \rightarrow 0$ limit
through (\ref{3+1}), that leads to the $\kappa$-Poincar\'{e} algebra.

The $SO_q(3,2)$ commutator on which we focus is~\cite{lukieIW}
\begin{eqnarray}
[M_{1,4},M_{2,4}] &=& {1 \over 2} {\sinh \left( z ( M_{12} + M_{04}) \right)
+ \sinh \left( z ( M_{12} - M_{04} ) \right)
\over \sinh (z)} \nonumber\\
& ~ & + {1 - e^{i z} \over 4 e^{ i z}} \left[ (i M_{03} -  M_{34})^2 - e^{i z}
(i M_{03} +  M_{34})^2 \right]
\label{so32qGeneric}
\end{eqnarray}
In the contraction the generators $M_{1,4}$, $M_{2,4}$,
$M_{34}$, play the role of the boosts $N_1$, $N_2$, $N_3$,
the generator $M_{1,2}$ plays the role of the rotation $M_3$
and the generators $M_{0,3}$ and $M_{0,4}$
are classically related to the $P_3$ and energy $E \equiv P_4$
by the Inonu-Wigner-contraction relation $P_\mu = \sqrt{\Lambda} M_{0,\mu}$.
However when taking the contraction in the quantum-gravity 3+1-dimensional
context we should renormalize according to (\ref{renormalize}),
and therefore
\f
P_{\mu ,ren} = \left ( {\sqrt{\Lambda} l_p \over \alpha } \right )^r
\sqrt{\Lambda} M_{0,\mu}
\label{renormalizebis}
\ff

Adopting (\ref{renormalizebis}), and taking into account that $z$
is given by (\ref{3+1}), one can easily verify that
the $\Lambda \rightarrow 0$ limit of (\ref{so32qGeneric})
is singular for $r>1$,
while for $r<1$ the limit is trivial and (\ref{so32qGeneric})
reproduces the corresponding commutator of the classical
Poincar\'{e} algebra. The interesting case is $r=1$,
where our framework indeed leads to the $\kappa$-Poincar\'{e}
algebra  ${\cal P}_{\kappa}(3+1)$
in the $\Lambda \rightarrow 0$ limit.
For $r=1$ and small $\Lambda$ the commutation relation (\ref{so32qGeneric})
takes the form
\begin{eqnarray}
[N_{1},N_{2}] &=& {\sinh \left( M_{3} l_p^2 \Lambda \right)
\over \sinh ( l_p^2 \Lambda)} \cosh \left( \alpha l_p E_{ren} \right)
+ \nonumber\\
& ~~ & + {l_p^2 \Lambda \over 4} \left[ \alpha^2 { P_{3,ren}^2 \over \Lambda}
- 2 \alpha { P_{3,ren} N_3 + N_3 P_{3,ren} \over l_p \Lambda}
+  i \alpha (P_{3,ren} N_3 + N_3 P_{3,ren}) \right]
\nonumber\\
& \rightarrow &  M_{3} \cosh \left( \alpha l_p E_{ren} \right)
+ {1 \over 4} \alpha^2 l_p^2 P_{3,ren}^2
- {1 \over 2} \alpha l_p [P_{3,ren} N_3 + N_3 P_{3,ren}]  .
\label{so32qGenericFINAL}
\end{eqnarray}

This indeed reproduces, for $\kappa = (\alpha l_p)^{-1}$,
the $[N_{1},N_{2}]$ commutator obtained in Ref.~\cite{lukieIW}
for the case in which the contraction of $SO_q(3,2)$
to ${\cal P}_{\kappa}(3+1)$ is achieved.
The reader can easily verify that for the other commutators
again the procedure goes analogously and in our framework, with $r=1$,
one obtains from $SO_q(3,2)$ the  full ${\cal P}_{\kappa}(3+1)$
described in Ref.~\cite{lukieIW}.

However, as we discussed in the 2+1-dimensional case,
the resulting presentation of the algebra suffers from the
problem that the boosts do not generate the ordinary lorentz
algebra. Hence,  we
must choose a different basis for
${\cal P}_{\kappa}(3+1)$ that does have a
Lorentz subalgebra.
In $3+1$ dimensions a basis that does have this property
was described by Majid-Ruegg in ~\cite{MajidRuegg}.
To arrive at the physical generators, we should
rewrite their basis in terms of renormalized energy-momentum
$E_{ren},P_{i,ren}$. One finds then a presentation of
the $\kappa$-Poincar\'{e} algebra, with an ordinary Lorentz
subalgebra. In this case the
deformed dispersion relation is given by,
\begin{eqnarray}
\cosh(\alpha l_p m) = \cosh(\alpha l_p E_{ren})
-\frac{\alpha^2 l_p^2}{2} \vec{P}_{ren}^2
e^{\alpha l_p E_{ren}}
~.
\label{disprel}
\end{eqnarray}

Having discussed the general structure of
the contraction $SO_q(3,2) \rightarrow {\cal P}_{\kappa}(3+1)$
which we envisage for the case of quantum gravity in 3+1 dimesions,
we turn to an analysis which is more specifically connected
to some of the results that recently emerged in the quantum-gravity
literature.
Specifically, we consider the boundary observables algebra
derived in Ref.~\cite{artem}. This is in fact (\ref{so31q}) with the
identifications (\ref{ep}), only here it is interpreted as the
algebra of the boundary observables in the $3+1$ dimensional theory.
However, now we want to take the limit appropriate to the
$3+1$ dimensional quantum deformation, (\ref{3+1}), and renormalize
according to (\ref{renormalize}).
From a technical perspective this requires us to repeat the
analysis of the previous Section (since the symmetry algebra
on the boundary of the 3+1-dimensional theory is again $SO_q(3,1)$
as for the bulk theory in 2+1 dimensions),
but adopting the renormalized energy-momentum
(\ref{renormalize}) and the relation (\ref{3+1}) between $q$ and $\Lambda$
which holds in the 3+1-dimensional case: $z = \ln q = l_p^2 \Lambda$.
The reader can easily verify that these two new elements
provided by the 3+1-dimensional context, compensate each other,
if $r=1$, and the contraction of $SO_q(3,1)$
proceeds just as in Section~2, leading again to ${\cal P}_{\kappa}(2+1)$.
In the context of the 3+1-dimensional theory we should see this
symmetry algebra as the projection of a larger 10-generator
symmetry algebra which,
in the limit $\Lambda \rightarrow 0$ and low energies, should
describe the symmetries of the ground
state. And indeed it is easy to recognize the
6-generator algebra ${\cal P}_{\kappa}(2+1)$
as the boundary projection of ${\cal P}_{\kappa}(3+1)$.

Thus, we reach similar conclusions to the $2+1$ case, with the
additional condition that in $3+1$ dimensions
the outcome of the contraction of
the symmetry algebra of the quantum theory depends on the
parameter $r$ that governs the renormalization of the
energy and momentum generators (\ref{renormalize}).
For the contraction to exist we must have $r \leq 1$.
For $r<1$ the contraction is the ordinary Poincar\'{e} algebra.
Only for the critical case of $r=1$ does a
deformed $\kappa$-Poincar\'{e} algebra emerge in the limit.

However, when this condition is satisfied, the conclusion that
the symmetry of the ground state is deformed is unavoidable,
the contraction must be some presentation of $\kappa$-Poincar\'{e}.
As in $2+1$ dimensions, the exact form of the algebra when expressed
in terms of the generators of physical symmetries cannot be determined
without additional physical input.  The algebra is restricted, but
not fixed, by the condition that it have an ordinary undeformed
Lorentz subalgebra. To fully fix the algebra requires the expression
of the generators of the low energy symmetries in terms of the
generators that define the symmetries of the full theory. These
presumably act on a boundary, as is described in \cite{artem}.

\section{Outlook}

Quantum gravity is a complicated subject, and the behavior of the low
energy limit is one of the trickiest parts of it. It has been argued
by many people recently, however, that quantum theories of gravity do
make falsifiable predictions, because they predict
modifications in the energy-momentum relations. The problem has
been how to extract the energy momentum relations reliably from the
full theory, and in particular to determine whether lorentz invariance
is broken, left alone, or deformed.

Here we have shown that the answers are in fact controlled by a
symmetry algebra, which constrains the theory and limits the possible
behaviors which result. Assuming only that the theory must be derived
as a limit of the theory with non-zero cosmological constant, we
have argued here that in $3+1$ dimensions
the symmetry of the ground state and the
resulting dispersion relation is determined partly by two parameters,
$r$ and $\alpha$, which arise in the renormalization of the hamiltonian,
(\ref{renormalize}).  The additional information required to determine
the energy-momentum relations involves an understanding of how the
generators of symmetries of the low energy theory are expressed in
terms of generators of symmetries of the full, non-perturbative
theory.

\section*{Acknowledgements}
We would like to thank  Laurent Freidel, Jerzy Kowalski-Glikman
and Joao Magueijo for conversations during the course of this
work.

\section*{Appendix: Quantum symmetry in $2+1$ dimensions.}
We summarize here one route to the quantization of $2+1$ gravity
which shows clearly the role of quantum symmetries, given by
Nelson and Regge\cite{regge}.

The route taken to quantize the theory is to solve the constraints
first and then apply quantization rules to the resulting reduced
phase space. As the connection $A^{ab}$ is flat by constraint
equations the reduced phase space is the moduli space of flat
connections modulo gauge transformations. This space can be
parameterized as the space of all homomorphisms from the
fundamental group of the surface $\Sigma$, $\pi_1(\Sigma)$, to the
gauge group. Such homomorphism can be realized by taking
holonomies of the connection  $A^{ab}$ along non-contractible
loops which arise due to handles of the surface $\Sigma$ and
punctures with particles inserted in them. The fundamental group
$\pi_1(\Sigma)$ thus depends on the genus of the surface $g$ and
the number of punctures with particles $N$, and consists of $2g+N$
generators $u_i$,$v_i$,$m_j$, where $i=1...g$, $j=1...N$. To each
of these generators should be associated an element of the gauge
group $U_i=\rho(u_i)$,$V_i=\rho(v_i)$,$M_j=\rho(m_i)$, satisfying
the following relation:
\begin{equation}
U_1V_1U_1^{-1}V_1^{-1}...U_gV_gU_g^{-1}V_g^{-1}M_1...M_N=1.
\end{equation}

The physical observables are now gauge invariant functions of
$U_i$,$V_i$,$M_j$, and the canonical commutation relations
(\ref{ccr}) define a poisson structure on the space of such
functions. In quantum theory the poisson brackets has to be
replaced by commutators and as a consequence the algebra of
functions on the gauge group representing the physical observables
becomes a noncommutative algebra. This can be understood as a
quantum deformation of the gauge group.

The detailed description of the poisson structure on the space of
functions of $U_i$,$V_i$, and$M_j$ can be found in \cite{fock}.
Here we will illustrate the origin of quantum group relations on a
simple example of two intersecting loops as it was first done in
\cite{regge}. Let $u$ and $v$ be two elements of the fundamental
group associated to the same handle (so that the corresponding
loops intersect). To each of them is associated an element of the
gauge group $U=\rho(u)$, $V=\rho(v)$. Given that $SO(2,2)\sim
SL(2,R) \oplus SL(2,R)$ and $SO(3,1) \sim SU(2) \oplus SU(2)^*$
each element can be decomposed as a sum of irreducible $2{\times} 2$
matrices $U=U^+ \oplus U^-$, $V=V^+ \oplus V^-$. The gauge
invariant functions that can be constructed from them are
$c^{\pm}(u)=TrU^{\pm}$, $c^{\pm}(v)=TrV^{\pm}$, and $c^{\pm}(uv)=Tr U^{\pm}
V^{\pm}$. In quantum theory they satisfy the following commutation
relation induced by (\ref{ccr})
\begin{eqnarray}
[c^{\pm}(u),c^{\pm}(v)]&=&{\pm}\frac{i\hbar 2 \pi}{k}\big(c^{\pm}(uv)-
c^{\pm}(uv^{-1})\big)={\pm}\frac{i\hbar 4 \pi}{k}\big(c^{\pm}(uv)-
c^{\pm}(u) c^{\pm}(v) \big), \nonumber \\
{[}c^{\pm} (u), c^\mp (v){]} &=& 0.
 \label{qcr}
\end{eqnarray}
For definiteness let us consider the case of negative cosmological
constant in which the gauge invariant functions defined above are
real and restrict ourselves to the '+' sector of the gauge group.
By introducing new variables $c^+(uv)=\sin \mu$,
$K^{\pm}=e^{iz/2}c^+(u) {\pm} i c^+(v) e^{{\pm} i \mu}$, where $2 \pi \hbar k^{-1}
= -2 \tan(z/2)$ the commutation relations (\ref{qcr}) can
be rewritten as  the following algebra
\begin{equation}
[\mu,K^{\pm}]={\pm} z K^{\pm}, \ \ [K^+,K^-]=\sin z \sin 2\mu.
\label{qalg0}
\end{equation}
This algebra up to rescaling coincides with the algebra
$SL_q(2,R)$. Analogously one can derive the algebra of functions
on the '-' sector of the gauge group which is also $SL_q(2,R)$. By
combining them together one finds that the gauge group in the case
of negative cosmological constant is $SO_q(2,2)$ and analogously
in the case of positive cosmological constant it is $SO_q(3,1)$.
%\end{multicols}{1}


\begin{thebibliography}{99}

\bibitem{qgprev} G.~Amelino-Camelia, gr-qc/991089,
{\it Are we at the dawn of quantum-gravity phenomenology?},
Lect.~Notes Phys.~541 (2000) 1;
T.~Kifune,
{\it Invariance violation extends the cosmic ray horizon?},
Astrophys.~J.~Lett.~{518}, L21 (1999);
R.J.~Protheroe and H.~Meyer,
{\it An infrared background TeV gamma ray crisis?},
Phys.~Lett.~{B493}, 1 (2000);
N.E.~Mavromatos, gr-qc/0009045;
S.~Sarkar,
{\it Possible Astrophysical Probes of Quantum Gravity},
%Invited talk at 1st IUCAA Workshop on Interface of Gravitational
%and Quantum Realms, Pune, India, 17-21 Dec 2001.
gr-qc/0204092;
D.V.~Ahluwalia,
{\it Interface of Gravitational and Quantum realms},
%Invited talk at 1st IUCAA Workshop on Interface of Gravitational
%and Quantum Realms, Pune, India, 17-21 Dec 2001.
gr-qc/0205121;
T.J.~Konopka, S.A.~Major, {\it Observational Limits on Quantum Geometry
Effects}, hep-ph/0201184, New J.~Phys.~4 (2002) 57; T.~Jacobson,
S.~Liberati and
D.~Mattingly, {\it TeV Astrophysics Constraints on Planck Scale
Lorentz Violation},
hep-ph/0112207;
R.C.~Myers and M.~Pospelov,
{\it  Ultraviolet modifications of dispersion relations in effective field
theory},
hep-ph/0301124;
R.H.~Brandenberger, J.~Martin,
 {\it  On Signatures of Short Distance Physics in the Cosmic Microwave
 Background}, hep-th/0202142, Int.~J.~Mod.~Phys.~A17 (2002) 3663.


\bibitem{dsr1} G.~Amelino-Camelia, {\it Relativity in space-times
with short-distance structure governed by an observer-independent
(Planckian) length scale},    gr-qc/0012051,
Int.~J.~Mod.~Phys.~D11 (2002) 35;
{\it Testable scenario for Relativity with minimum length}, hep-th/0012238,
Phys.~Lett.~B510 (2001) 255.

\bibitem{dsrpaps} J.~Kowalski-Glikman, {\it Observer
Independent Quantum of Mass},
hep-th/0102098, Phys.~Lett.~A286 (2001) 391;
N.R.~Bruno, G.~Amelino-Camelia and J.~Kowalski-Glikman,
{\it Deformed boost transformations that saturate at the Planck length},
hep-th/0107039, Phys.~Lett.~B522 (2001) 133;
J.~Magueijo and L.~Smolin, {\it Lorentz invariance with
an invariant energy scale}, hep-th/0112090, Phys.~Rev.~Lett.~88 (2002) 190403;
S.~Judes and M.~Visser, {\it Conservation Laws in Doubly Special Relativity}
gr-qc/0205067.

\bibitem{various} J.~Kowalski-Glikman and S.~Nowak,
{\it Noncommutative space-time of doubly special relativity theories},
hep-th/0204245,
Int.~J.~Mod.~Phys.~D12 (2003) 299;
J.~Magueijo and L.~Smolin,{\it Generalized lorentz invariance
with an invariant energy scale}, gr-qc/0207085, Phys.Rev. D67 (2003) 044017;
{\it Gravity's rainbow} gr-qc/0305055;
G.~Amelino-Camelia, {\it Doubly Special Relativity}, gr-qc/0207049,
Nature 418 (2002) 34;
{\it Doubly Special Relativity: First results and key open problems},
gr-qc/0210063,
Int.~J.~Mod.~Phys.~D11 (2002) 1643;
D.~Kimberly, J.~Magueijo, J.~Medeiros,
{\it Non-Linear Relativity in Position Space},
gr-qc/0303067.

\bibitem{gampulmexweave} R.~Gambini and J.~Pullin,
{\it Nonstandard optics from quantum spacetime}, gr-qc/9809038,
Phys.~Rev.~D59 (1999) 124021; J.~Alfaro,
H.A.~Morales-Tecotl and L.F.~Urrutia,
{\it Loop quantum gravity and light propagation},
hep-th/0108061, Phys.~Rev.~D65 (2002) 103509;
{\it Quantum Gravity corrections to neutrino propagation},
gr-qc/9909079
Phys.~Rev.~Lett.~84 (2000) 2318.

\bibitem{grbgac} G.~Amelino-Camelia, J.~Ellis,
N.E.~Mavromatos, D.V.~Nanopoulos and S.~Sarkar,
astro-ph/9712103, Nature 393 (1998) 763.

\bibitem{jacob} G.~Lambiase, Gen.~Rel.~Grav.~34 (2002) 1437.

\bibitem{positive} L.~Smolin, {\it Quantum gravity with a
positive cosmological constant}, hep-th/0209079.

\bibitem{triangles} J.~Ambjorn,
{\it Quantum Gravity Represented As Dynamical Triangulations},
Class. Quant. Grav. 12 (1995) 2079; Agishtein and A. Migdal,
  {\it Simulations Of Four-Dimensional Simplicial Quantum Gravity},
 Mod.Phys.Lett. A7 (1992) 1039;
J. Ambjorn, A. Dasgupta, J. Jurkiewiczcy and R. Loll,
{\it A Lorentzian cure for Euclidean
troubles}, hep-th/0201104; J. Ambjorn and R. Loll,
{\it Non-perturbative Lorentzian Quantum Gravity,
Causality and Topology Change},
hep-th/9805108, Nucl.~Phys.~B536 (1998) 407.

\bibitem{reggecalc} H. Hamber and R.M. Williams,
%{\it Higher Derivative Quantum Gravity on a Simplicial Lattice},
Nuclear Physics B248 (1984) 392;  B260 (1985) 747; B269 (1986) 712;
 B267 (1986) 482; B400 (1993) 347.
% {\it  Phases of Simplicial Quantum Gravity in Four Dimensions: Estimates
%for the Critical Exponents},
%Nuclear Physics B400, 347-389 (1993).

\bibitem{kodama} H.~Kodama, {\it Holomorphic Wave Function Of The
Universe},  Phys.~Rev.~D42 (1990) 2548; L.~Smolin, Chopin Soo,
{\it The Chern-Simons Invariant as the Natural Time
Variable for Classical and Quantum Cosmology},
gr-qc/9405015,  Nucl.~Phys.~B449 (1995) 289.

\bibitem{witten} A.~Achucarro and P.~Townsend, {\it A
Chern-Simons Action For Three-Dimensional Anti-De Sitter Supergravity
Theories},  Phys.Lett. B180 (1986) 89; E. Witten, {\it (2+1)-Dimensional
Gravity As An Exactly Soluble System},  Nucl.~Phys.~B311 (1988) 46.

\bibitem{regge} J.E. Nelson, T. Regge, F. Zertuche,
Nucl. Phys. B339 (1990) 316.

\bibitem{NR1} E.~Buffenoir, K.~Noui, P.~Roche,
{\it Hamiltonian Quantization of Chern-Simons theory with SL(2,C)
Group},
hep-th/0202121,  Class.Quant.Grav. 19 (2002) 4953.

\bibitem{linking} L.~Smolin,
{\it Linking Topological Quantum Field Theory and Nonperturbative Quantum
Gravity},gr-qc/9505028, J.~Math.~Phys.~36 (1995) 6417.

\bibitem{baez-deform} J.C.~Baez,
{\it An Introduction to Spin Foam Models of Quantum Gravity and BF Theory},
gr-qc/9905087,  Lect.~Notes Phys.~543 (2000) 25.

\bibitem{qdef} S.~Major, L.~Smolin,
{\it Quantum deformation of quantum gravity},
 gr-qc/9512020, Nucl.Phys. B473 (1996) 267.

\bibitem{artem} A.~Starodubtsev, {\it Topological excitations around
the vacuum of quantum gravity I: the symmetries of the vacuum},
hep-th/0306135.

\bibitem{lukieAnnPhys} J.~Lukierski, A.~Nowicki and H.~Ruegg,
{\it Classical and quantum-mechanics of free $\kappa$-relativistic systems},
Ann. Phys. {243} (1995) 90-116.

\bibitem{hooft} S. Deser, R. Jackiw, and G. t'Hooft,
Ann. Phys. 152 (1984) 220.

\bibitem{inWig} E.~Inonu and E.P.~Wigner,
Proc.~Natl.~Acad.~Sci.~U.S.~39 (1953) 510.

\bibitem{lukieDdim} J.~Lukierski and H.~Ruegg,
{\it Quantum kappa Poincare in any dimension},
hep-th/9310117,
Phys.~Lett.~B329 (1994) 189.

\bibitem{firenze} E.~Celeghini, R.~Giacchetti, E.~Sorace and M.~Tarlini,
J.~Math.~Phys.~31 (1990) 2548.

\bibitem{lukieIW} J.~Lukierski, A.~Nowicki and H.~Ruegg,
Phys.~Lett.~B293 (1992) 344.

\bibitem{MajidRuegg} S.~Majid and H.~Ruegg, hep-th/9405107,
Phys.~Lett.~B334 (1994) 348.

\bibitem{leeNO2}A. Ashtekar, J. Baez, K. Krasnov,
{\it Quantum Geometry of Isolated Horizons and Black Hole Entropy},
gr-qc/0005126,  Adv.Theor.Math.Phys. 4 (2000) 1-94.

\bibitem{leeNO3} L. Freidel, K. Krasnov, R. Puzio,
{\it  BF Description of Higher-Dimensional Gravity Theories},
hep-th/9901069, Adv.~Theor.~Math.~Phys. 3 (1999) 1289.

\bibitem{leeNO4} K.~Ezawa,
{\it Ashtekar's formulation for $N=1,2$ supergravities as "constrained"
BF theories},
hep-th/9511047,  Prog.~Theor.~Phys. 95 (1996) 863;
Y.~Ling, L.~Smolin, {\it Eleven dimensional
supergravity as a constrained topological field
theory}, hep-th/0003285,  Nucl.~Phys.~B601 (2001) 191.

\bibitem{leeNO7} J.F.~Plebanski, {\it On
the separation of einsteinian substructures},
J. Math. Phys., 18 (1977) 2511;
R.~Capovilla, J.~Dell and T.~Jacobson, Phys. Rev. Lett. 21 (1989) 2325;
Class. Quant. Grav. 8 (1991) 59; R. Capovilla, J. Dell, T. Jacobson and
L. Mason, Class. and Quant. Grav. 8 (1991) 41.

\bibitem{leeNO8} J.W.~Barrett, L.~Crane,
{\it Relativistic spin networks and quantum gravity},
gr-qc/9709028,  J.~Math.~Phys. 39 (1998) 3296.

\bibitem{NR2} K.~Noui, P.~Roche, {\it Cosmological
Deformation of Lorentzian Spin Foam Models},
gr-qc/0211109.

\bibitem{leeNO9} L.~Smolin, {\it A
holographic formulation of quantum general relativity},
hep-th/9808191,  Phys.~Rev. D61 (2000) 084007;
Y.~Ling, L.~Smolin, {\it Holographic
Formulation of Quantum Supergravity},
 hep-th/0009018,  Phys.~Rev.~D63 (2001) 064010.

\bibitem{CY} L. Crane, D.N. Yetter,
{\it Measurable Categories and 2-Groups},
math.QA/0305176.

\bibitem{fock}  V.V.~Fock, A.A.~Rosly,
Am.~Math.~Soc.~Transl.~191 (1999) 67,  math.QA/9802054.

\end{thebibliography}
\end{document}